\documentclass[amsmath,aps,twocolumn, superscriptaddress,letterpaper, amsmath,amssymb
]{revtex4-1}

\usepackage{graphicx}
\usepackage{dcolumn}
\usepackage{bm}
\usepackage{color}
\usepackage{appendix}

\newcommand{\ii}{\textmd{i}}

\newcommand{\dd}{\textmd{d}}
\newcommand{\tup}{\tau_\mathrm{up}}
\newcommand{\tdo}{\tau_\mathrm{dn}}
\newcommand{\td}{\tau_\mathrm{d}}

\newcommand{\avgvyi}{\langle \mathbf{v}(y_\textmd{i}) \rangle} 
\newcommand{\be}{\begin{equation}}
\newcommand{\ee}{\end{equation}}

\newcommand{\PLH}{{\mkern-2mu\times\mkern-2mu}}

\begin{document}

\title{Intermittent Granular Dynamics at a Seismogenic Plate Boundary}

\author{Yasmine Meroz}
\email{ymeroz@seas.harvard.edu}
\affiliation{John A. Paulson School of Engineering and Applied Sciences,  Harvard University, Cambridge MA, 02138, USA.}

\author{Brendan J. Meade}%
\affiliation{Department of Earth and Planetary Sciences, Harvard University, Cambridge MA, 02138, USA}%

\date{\today}

\begin{abstract}
Earthquakes at seismogenic plate boundaries are a response to the differential motions of tectonic blocks embedded within a geometrically complex network of branching and coalescing faults.
 Elastic strain is accumulated at a slow strain rate of the order of $10^{-15}$ s$^{-1}$, and  released intermittently at intervals $>100$ years, in the form of rapid (seconds to minutes) coseismic ruptures.
The development of macroscopic models of quasi-static planar tectonic dynamics at 
these plate boundaries has remained challenging due to uncertainty with regard to the spatial and kinematic complexity of fault system behaviors. In particular, the characteristic length scale of kinematically distinct tectonic structures is poorly constrained. 
Here we analyze fluctuations in GPS recordings of interseismic velocities from the southern California plate boundary, identifying heavy-tailed scaling behavior. This suggests that the  plate boundary can be understood as a densely packed granular medium near the jamming transition, with a characteristic length scale of $91 \pm 20$ km.
In this picture fault and blocks systems may rapidly rearrange the distribution of forces within them, driving a mixture of transient and intermittent fault slip behaviors over tectonic time scales.

\end{abstract}

\maketitle

Fault slip at seismogenic plate boundaries is rarely continuous, instead occurring intermittently during short duration earthquakes which last from seconds~\cite{abercrombie1995earthquake} to minutes~\cite{ishii2005extent} during which fault slip may reach up to 50 meters~\cite{ozawa2011coseismic}. Throughout the interseismic interval between large earthquakes, plate boundaries are not dormant but instead slowly accumulate the elastic strain that will be released in future earthquakes~\cite{lawson1908california}, while the faults themselves are frictionally locked. In addition to this temporally bi-modal behavior, fault slip at plate boundaries is not localized along a single fault but rather spread across fault systems where faults are bounded by tectonic blocks to form anastomosing fault systems~\cite{Plesch2007, 
jennings1975fault, wallace2004subduction}.  



\begin{figure}[t]
\centering
\includegraphics[width=0.98\linewidth]{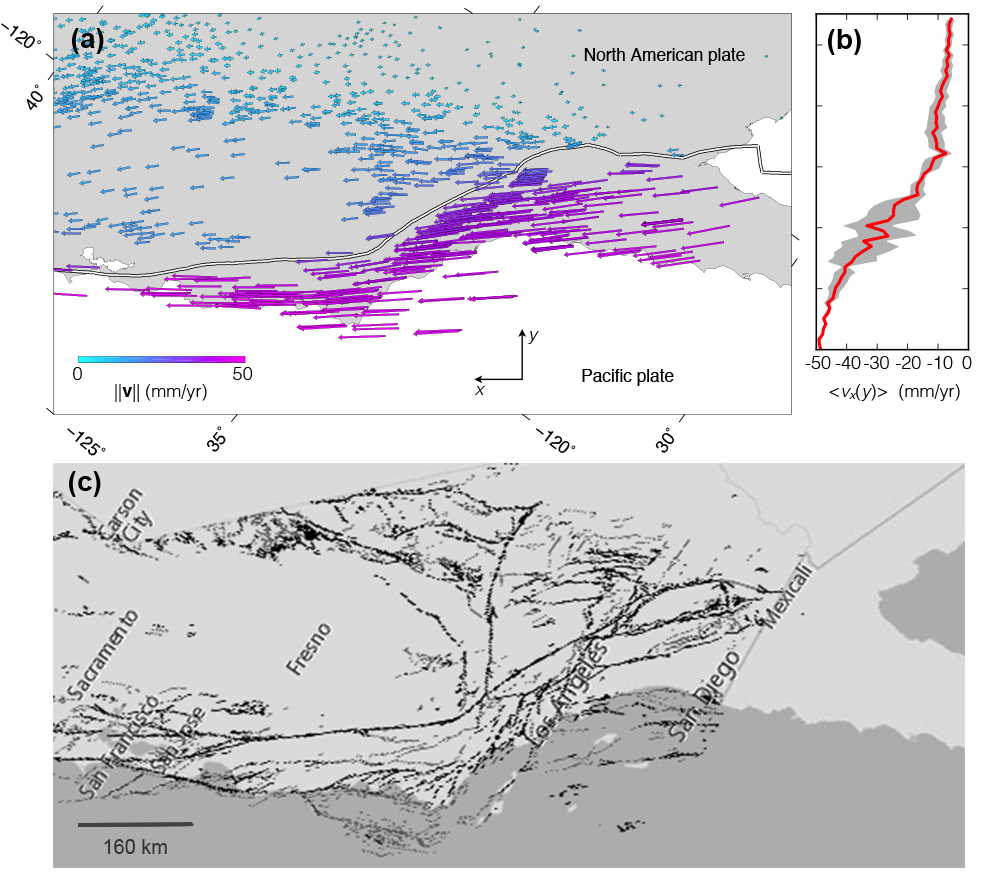}
\caption{
(a) Vector field of $1,106$ interseismic GPS velocities from across the Pacific-North American plate boundary in California~\cite{kreemer2014geodetic}, where magnitude is given by color and size. The x-axis is approximately parallel to the trace of the  San Andreas fault shown here, while the y-axis runs across $L=565$ km from the Pacific to the North American plate. 
(b) Average velocity profile 
as a function of y position in (a). Standard deviation in grey. 
(c) The full fault system: a geometrically complex network of branching and coalescing faults which accommodate the deformation between the Pacific and North American plates~\cite{fault_activity_online}, spread across $>100$km. 
}
\label{fig:vels}
\end{figure}

Dynamic models of activity at seismogenic plate boundaries have generally focused on the physics within narrow ($<1$m) individual fault shear zones~\cite{fisher1997statistics, dahmen1998gutenberg, aharonov1999rigidity, aharonov2004stick, bouchbinder2013}. However while large faults (e.g. the San Andreas fault in California)  accommodate the majority of deformation between two plates (e.g. the Pacific and North American plates), they are embedded within a geometrically complex network of branching and coalescing faults~\cite{fault_activity, Plesch2007}, as can be appreciated in  Fig.~\ref{fig:vels}, and the interactive map in~\cite{fault_activity_online}.

Here we investigate the macroscopic dynamics of planar tectonics within such fault and block systems occurring on scales $>100$km, while considering the suggestion~\cite{anderson2007new} that sheared granular systems may provide an analog.
Both systems exhibit intermittent dynamics manifested in a gradual increase in elastic energy  released by a precipitous event; in slowly sheared granular media these are rearrangement events of single granules, while in seismogenic plate boundaries these are earthquakes.
Moreover in both systems the motion at the boundaries is defined and results from the differential motion of the granules or tectonic blocks adjoining a potential slip surface. Here we consider the extent to which scaling laws derived from numerical simulations of slowly sheared 2D deformable granular systems (foam)~\cite{Ono2003, Durian1995, Durian1997}, may describe earthquake cycle activity at tectonic plate boundaries, including spatial variabilities of nominally interseismic GPS velocities across the Pacific-North America plate boundary in California, as well as characteristic earthquake cycle time scales. 


We consider $1,106$ nominally interseismic GPS velocity observations, recorded over a period of $\sim 10$ years on average, in the area sheared between the Pacific and North American plates in California \citep{kreemer2014geodetic}. The area is defined as running from the Pacific plate at the bottom, where we define $y=0$, to the North American plate at the top where $y= L = 565$ km, while the $x$-axis runs along $0 < x < 1,200$ km, approximately parallel to the San Andreas fault.
The velocities $\mathbf{v}_\ii$ at each GPS reading i are calculated relative to the North American plate, and are shown in Fig.~\ref{fig:vels}a, represented by a vector field. Relative velocities exhibit a clear trend, with lower absolute values closer to the North American plate (here considered fixed), and increasing values closer to the moving Pacific plate, with a maximum of $\sim 50$ mm/year.
We calculate the average velocity profile along the $y$-axis, $\langle \mathbf{v}(y) \rangle$, by  discretizing the $y$-axis into $N=65$ bins, with $\dd y = 8.7$ km. 
The discretized velocity profile is then calculated as the average velocity of readings within the same bin. 
For simplicity in what follows we will adopt
the continuous notation $\langle \mathbf{v}(y) \rangle$. The average velocity profile is shown in Fig.~\ref{fig:vels}b,  exhibiting non-linearity, i.e. shear bands, as also observed experimentally in different sheared granular systems ~\cite{Lauridsen2002}. The shear strain rate, defined as the relative velocity of the two plates divided by the distance between them, i.e. $\dot{\epsilon} = (\langle v_x(L) \rangle - \langle v_x(0) \rangle )/ L$, yields $\dot{\epsilon} = 7.66\mathrm \times 10^{-8}$ (1/yr).
We define the fluctuation of the velocity of a reading from the average velocity profile as:
\be\label{eq:dvi}
\delta \mathbf{v}_\ii = \mathbf{v}_\ii - \avgvyi ,
\ee
where $\mathbf{v}_\ii$ is the velocity of reading i, and $\avgvyi$ is the average velocity at position $y_\ii$ associated with reading i. 
To characterize the kinematics of the system as manifested by the velocity fluctuations, we calculate the distribution of velocity fluctuations of the GPS readings $P(\delta\mathbf{v})$, and the equal-time spatial correlation function 
\be
\label{eq:Fr}
C(r) = \frac{\langle \delta\mathbf{v}(r) \cdot \delta\mathbf{v}(0) \rangle }{ \langle \delta\mathbf{v} ^2\rangle}. 
\ee
We consider the analogy between the intermittent earthquake events, which release energy accumulated due to the shearing of tectonic plates along a fault, and the intermittent rearrangement events in sheared 2D granular systems, and compare our results to those of numerical simulations of slowly sheared 2D deformable granular medium (foam)~\cite{Ono2003}. For slow shearing rates the distribution of velocity fluctuations has been found to deviate from a Gaussian (expected for fluid-like flow at high shearing rates), and spatial correlations follow a stretched exponential (deviating from the exponential expected for fluid behavior).

\begin{figure}[t]
\centering
\includegraphics[width=\linewidth]{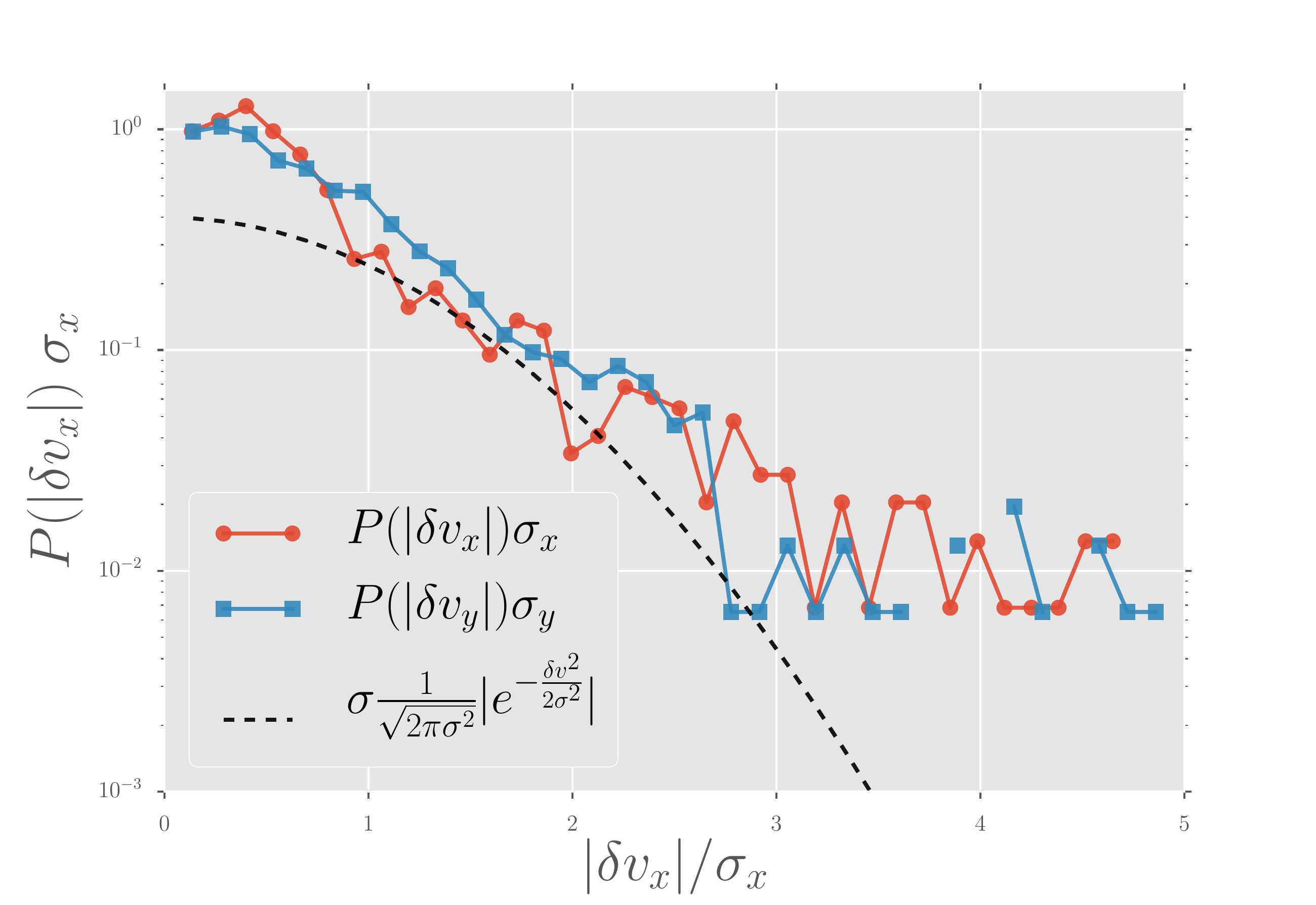}
\caption{Plot of the distributions of the velocity fluctuations $P(\Delta v_x)$ (red circles) and $P(\Delta v_y)$ (blue squares). We collapse the distributions by plotting them  multiplied by the standard deviation $\sigma_x$ or $\sigma_y$ accordingly, as a function of the absolute value of velocity fluctuations re-scaled by the standard deviation~\cite{Ono2003}. We use a semi-logarithmic scale. A Gaussian distribution (dashed line) is plotted for comparison, highlighting that heavy tails of the distributions, which clearly deviate from a Gaussian at high velocities.}
\label{fig:Pv}
\end{figure}

The distribution of velocity fluctuations of the GPS readings in the x and y directions , $P(\delta v_x)$ and $P(\delta v_y)$, are plotted in Fig.~\ref{fig:Pv}, and their fit to a Gaussian distribution is added for comparison. The distributions deviate significantly from a Gaussian distribution, exhibiting heavy tails. 
This is further confirmed by calculating the non-Gaussian parameter (NGP)~\cite{Rahman1964} for different moment ratios:
\be
\label{eq:ngp}
\alpha_n(x) = \frac{\langle x^{2n} \rangle}{ C_n\langle x^2 \rangle ^n} - 1,
\ee
where $C_n$ is a known constant. By definition, $\alpha_n$ equals zero for Gaussian distributions.  Here we calculate $\alpha_2$ and $\alpha_3$, where in 2D $C_2=3$ and $C_3=15$.
The NGP calculated for the fluctuations in the x direction yield  $\alpha_2 = 0.25$ and $\alpha_3 = 0.47$, and in the $y$ direction $\alpha_2 = 0.62$ and $\alpha_3 = 2.18$. For comparison, we calculate these values for a numerically generated Gaussian distribution with an identical population size, yielding $\alpha_2 = -0.05$ and $\alpha_3 = -0.18$, highlighting that the NGP values calculated for the original distributions clearly indicate non-Gaussianity.

We note that a part of the San Andreas fault deviates from the horizontal orientation of the plates within the area  analyzed here, as depicted in Fig~\ref{fig:vels}a. In order to rule out the possibility that the broad distribution of velocity fluctuations may be an artifact due to this slope, we carry out the analysis for the left third of the plate boundary, where the fault is relatively horizontal (details in the Supporting Information - SI). We find that the heavy tails are even more pronounced (see Fig.~S2 in the SI), showing the independence of our results on the geometric form of the San Andreas fault. Moreover, substantial fluctuations are not limited to the San Andreas fault, but rather are exhibited in an area of $\sim 100$km across the fault (see Fig.~S1 in the SI).

\begin{figure}[t]
\centering
\includegraphics[width=\linewidth]{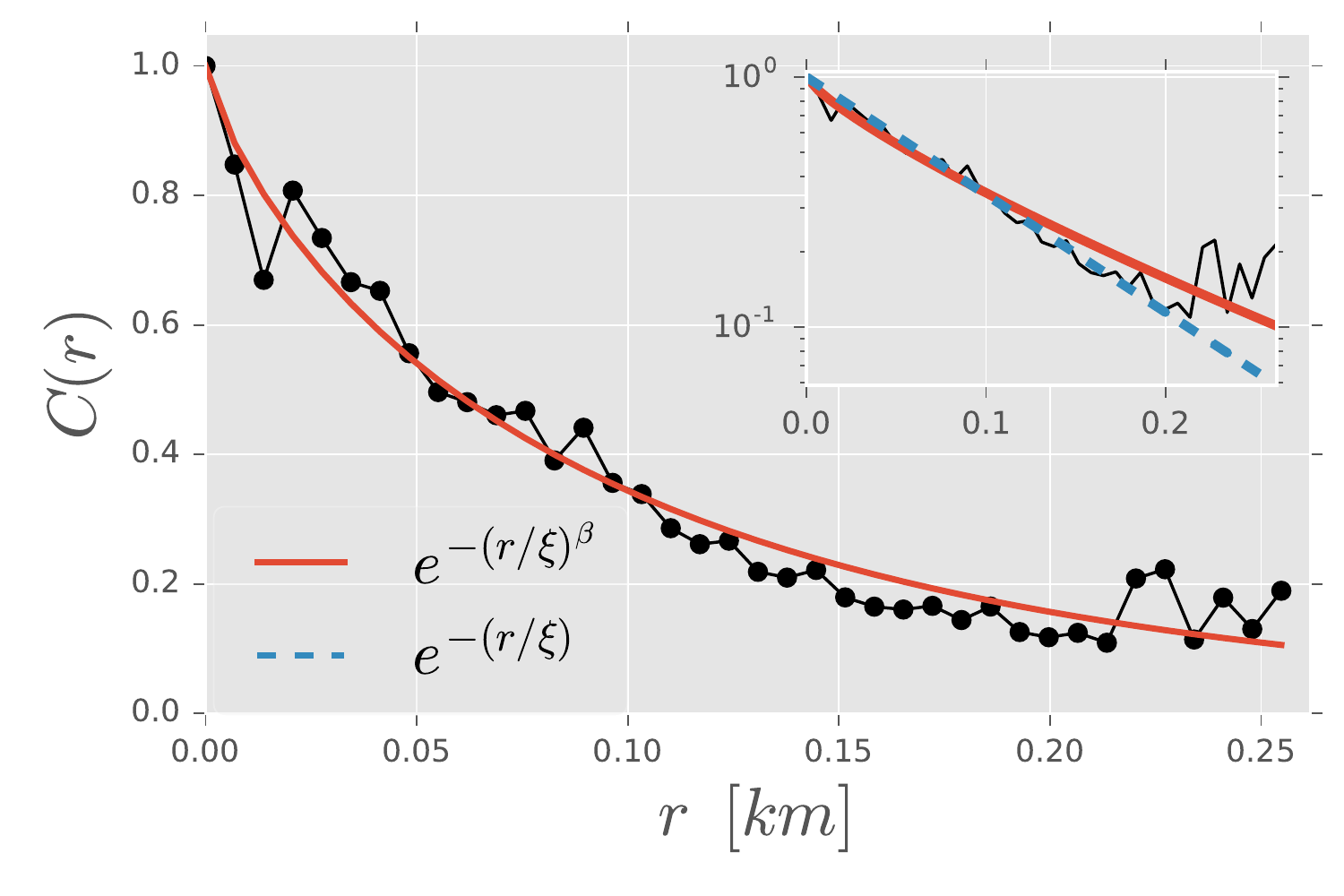}
\caption{Equal-time spatial correlation function of velocities fluctuations calculated as in Eq.~\ref{eq:Fr}, $C(r) = \langle \delta\mathbf{v}(r) \cdot \delta\mathbf{v}(0) \rangle / \langle \delta\mathbf{v} ^2\rangle$  (dotted line), and fit to the stretched exponential function in Eq.~\ref{eq:Fr_stretch}, $e^{-(r/\xi)^{\beta}}$ (solid line). The fit yields $\beta=0.75$, and $\xi=92$ km, with $R^2=0.92$ of the least squares fit. The inset plots the correlation function on a semilog scale, with an exponential (dashed blue) for comparison.}
\label{fig:Fr_fit}
\end{figure}

To calculate the spatial distribution of velocity fluctuations we  limit ourselves to  $\sim 200$km around the fault (roughly $100\mathrm{km} < y <300 \mathrm{km}$), thus avoiding recordings associated with the plates themselves which would mask the relevant correlation.
In Fig.~\ref{fig:Fr_fit} we plot the calculated spatial correlation of velocities fluctuations according to Eq.~\ref{eq:Fr}.
We fit this to the general form of a stretched exponential:
\be
\label{eq:Fr_stretch}
F(r) = e^{-(r/\xi)^{\beta}},
\ee
yielding the values $\beta=0.75$, and $\xi=92$ km, with $R^2=0.92$ of the least squares fit.  Simulations for 2D sheared foam at low shearing rates~\cite{Ono2003} exhibit values of $\beta < 1$, deviating from a regular exponential behavior with $\beta=1$, and $\xi$ is found to be associated with the average granule or bubble size.

\begin{table}[t]
\centering
\begin{tabular}{lll}
\hline
Parameter & Description & Value  \\
\hline
$\tdo$ & duration of earthquake & 10-600 s \citep{scholz2002mechanics}\\
$\tup$ & earthquake recurrence interval  & $10^{8}-10^{11.5}$ s \citep{scholz2002mechanics}\\
$\dot{\epsilon}$ & shear strain rate & $10^{-15}$  $\mathrm{s}^{-1}$ \citep{shen2007implications}\\
$\tau_{\mathrm{d}}$ & crack tip evolution & $10^{-9}$ s \citep{ben2008collective}\\
$\dot{\gamma} = \dot{\epsilon} \tau_\mathrm{d}$ & normalized shear strain rate & $8\PLH 10^{-24}$  \\
\hline
$\tdo/\tup$ & observed ratio & $6\PLH 10^{-6}-3\PLH 10^{-11}$  \\
$ t_{\mathrm{dn}}^{pred} /  t_{\mathrm{up}}^{pred}$ & predicted ratio (eq.~\ref{eq:up_down_ratio}) & $5 \PLH 10^{-9}$ \\
\hline
\end{tabular}
\caption{ Observed and predicted~\cite{Ono2003} ratio of earthquake cycle timing $\tdo / \tup$, where $\tdo$ is the duration of an earthquake, equivalent to the time of a rearrangement event in a granular system, and $\tup$ is the time between earthquakes where the elastic strain is accumulated, equivalent to the time in between rearrangement events. The predicted ratio is within the range of observed values. Also shown are observed values for the shear strain rate  $\dot{\epsilon}$, the crack tip evolution $\tau_{\mathrm{d}}$, and the normalized shear strain rate $\dot{\gamma}$ , substituted in Eq.~\ref{eq:up_down_ratio} to yield the predicted ratio of earthquake cycle timing. Further details in the main text.}
\label{tab:earthquake_cycle_timing}
\end{table}


In this work we investigate macroscopic planar tectonic dynamics across a geometrically complex network of faults and blocks accommodating the deformation between the North American and Pacific plates in California on scales $>100$km. 
Here we consider the similarity between fault block systems and the intermittent dynamics of granular systems sheared at low shearing rates, both of which exhibit a volatile elastic energy over time. The elastic energy increases gradually as granules slowly deform (as blocks deform during the interseismic phase of the earthquake cycle), and decreases rapidly due to intermittent granule rearrangement events (due to earthquakes). 
Pursuing this analogy we find that the dynamics displayed by the readings of the seismogenic plate boundary, as manifested by their velocity fluctuations, are consistent with those characterizing  analogous 2D sheared deformable granular media, where the velocities $\mathbf{v}_\ii$ are associated with densely packed granules or bubbles constrained between two shearing plates at a low shearing rate.  
Namely we show that, consistent with findings for slowly sheared granular media~\cite{Ono2003}, the distribution of velocity fluctuations  $P(\delta\mathbf{v})$ deviates from a Gaussian, exhibiting broad tails, and the correlation function decays as a stretched exponential.
We note that the velocity profile deviates from the linear profile expected for a homogeneously sheared elastic body, ruling this out as an alternative mechanism.


This analogy, based on kinematic characteristics, allows to deduce the dynamics of the system, and to make predictions concerning time-scales and length-scales associated with the dynamics at a seismogenic plate boundary. 
One of the primary observations in fault block systems is the ratio of time needed for the coseismic release of elastic strain (earthquakes) and the time it takes for slow plate motions (10-100 mm/yr) to accumulate elastic strain.
Following the analogy with sheared granular systems~\cite{Ono2003}, we consider the timescales of the systems, 
defining $\tup$ as the average duration of the energy buildup in between intermittent sharp energy releases, and $\tdo$ as the average duration of these energy drops. 
Ono et al.~\cite{Ono2003} find that the ratio of these characteristic times follows
\begin{equation}
\label{eq:up_down_ratio}
\tdo / \tup = 7.9 \dot{\gamma}^{0.4},
\end{equation}
where $\dot{\gamma} = \dot{\epsilon} \td$ is the normalized shear rate, and $\td$ is the characteristic time scale in the model, the duration of a rearrangement event
~\cite{Durian1995,Ono2003}. 
%
%
For the earthquake system, $\tdo$ is the characteristic duration of an earthquake, and $\tup$ is the earthquake recurrence interval. Observed values and their ratio $\tdo/\tup$ are given in Table~\ref{tab:earthquake_cycle_timing}. 
The ratio is found to be consistent with the ratio predicted by Eq.~\ref{eq:up_down_ratio}, thus relating these timescales to the observed sheared strain rate $\dot{\epsilon}^{\mathrm{obs}}$, and the timescale of the system $\td$, suggested to be the nano-second value for the crack tip evolution time scale~\cite{ben2008collective}.

Moreover, we recall that in slowly sheared granular material the value of the stretched exponential fit of the spatial correlation function in Eq.~\ref{eq:Fr_stretch}, $\xi$,  
is associated with the average granule or bubble size~\cite{Ono2003}, suggesting that the length scale associated with the seismogenic plate boundary is roughly $90$ km. 
This value is 4-5 orders of magnitude larger than granular fault gauge \citep{aharonov2004stick}, indicating that plate boundaries may be treated as granular systems at a macroscopic scale with the length scale $\xi$ associated with the average mesh size of the fault network. 
Furthermore, we show that the heavy tails of the fluctuation distributions do not depend on the form of major fault accommodating most of the deformation between the two plates - the San Andreas fault (details in the SI). This further strengthens the  concept that the macroscopic tectonic dynamics are governed by the intricate fault and block system, rather than just the major fault.

Lastly, this model provides an explanation for the clustered and intermittent fault behaviors observed in the geologic record as a response to the rapid reorganization of force chains by earthquakes themselves. Geological observations of macroscopic fault system activity have revealed an array of behaviors, for example at the seismically active southern California plate boundary observations of paleo-earthquake activity range from nearly periodic \citep{fumal2002evidence, scharer2007paleoearthquakes}, to clustered \citep{grant1994paleoseismic}, and out of phase \citep{dolan2007long}. These observations of diverse fault activity can be explained by a model where plate boundaries are considered as macroscopic granular shear zones near the jamming transition \citep{ben2008collective} with effective granule sizes $>10$ km.  Granule sizes at this scale enable earthquakes themselves to redistribute forces within plate boundaries by creating and destroying force chain paths and producing complex time evolving fault slip rate histories.

\section*{Acknowledgments}
The Authors thank Ido Regev and Yohai Bar-Sinai for  helpful observations and fruitful conversations.




\bibliography{biblio}

\end{document}